\documentclass[%
reprint,
amsmath,amssymb,
aps,
floatfix,
]{revtex4-2}

\usepackage{graphicx}
\usepackage{wrapfig}
\usepackage{dcolumn}
\usepackage{bm}

\begin{document}


\title{Radio Astronomy with the Arecibo 305m Telescope: In Contemporaneous Context}

\author{Tapasi Ghosh}
\email{tapasig91@gmail.com}
\affiliation{Arecibo, and Green Bank Observatory (retired)}
\author{Chris Salter}
\email{csalter.wfc@gmail.com}
\affiliation{Arecibo Observatory (retired)}




\date{\today}

\begin{abstract}
Most scientific research begins in the context of the then contemporary state of knowledge of the field and moves towards a deeper understanding of the subject. In this colloquium, we present the Arecibo telescope's contribution to radio astronomy from the point of view of its contemporary impact and relevance, and how that  evolved over the 57 years of its long life. Sometimes, serendipitous discoveries at Arecibo and elsewhere brought revolutionary changes to the field. Further, significant upgrades to Arecibo's reflector, optics, receiving and data-taking systems helped clear the pathway to leap ahead. Charting these movements through time, and without any claim to completeness, this Colloquium aims to present a progression through Arecibo's role in the history of radio astronomy.
 
\end{abstract}

\maketitle

\tableofcontents
\section{Introduction: The Dawn, 1963 - 67}
When the giant 305-m dish saw first light in late 1963 \cite{gordon64}, 
\begin{figure}
\includegraphics[width=0.45\textwidth]{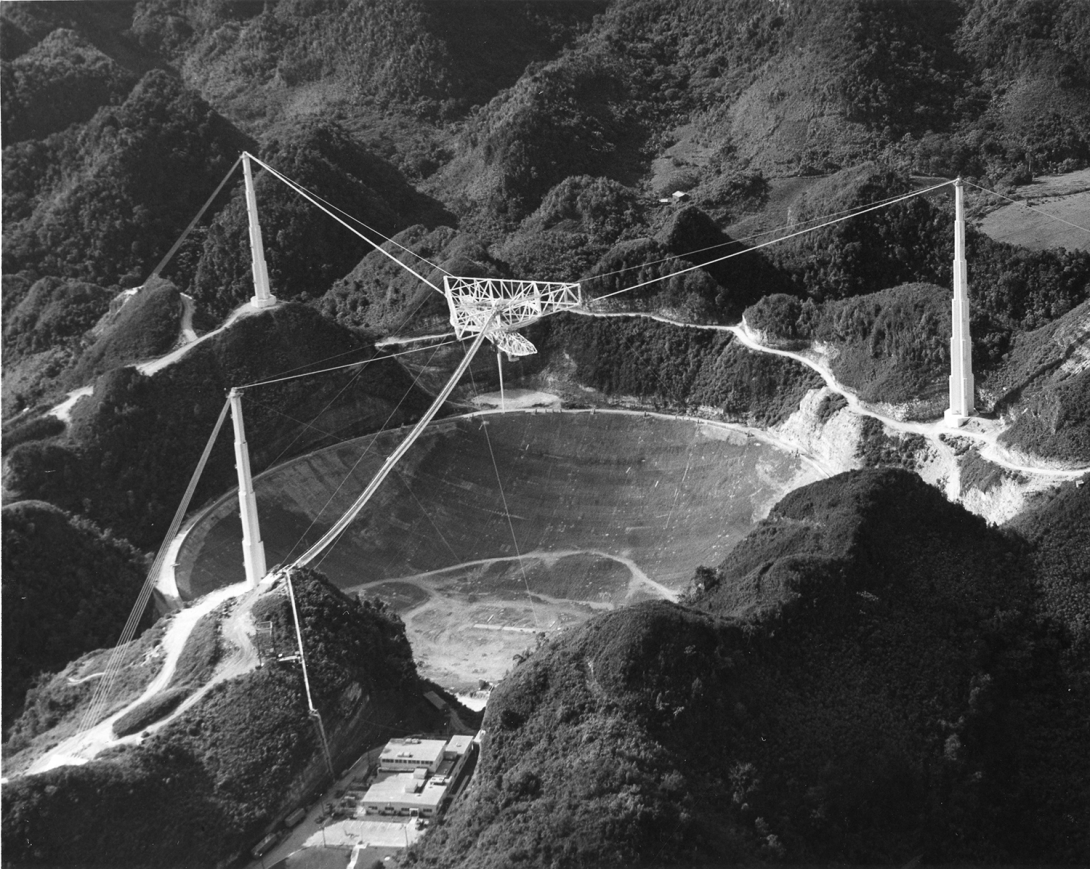}
\caption{\label{fig:ao63}Arecibo telescope at first light in 1963. Courtesy of
Cornell University and AIP Emilio Segr`e Visual Archives.}
\label{fig:ao_1963}
\end{figure}
the most important astronomy ``news of the day'' was the optical identification of 3C273 with a new class of objects, called QUASARs; Quasi Stellar Radio Sources \cite{schmidt1963}. The Radio Sky at that time was known to be dominated  by the Galactic continuum radiation emitted  via the synchrotron mechanism by high-energy cosmic-ray electrons, and many discrete radio sources whose nature astronomers were trying to decipher. The pathway for this was to determine their positions with an accuracy of $\sim$\,1\,arcsec, enabling their identification with optical objects. This led to obtaining their distances via redshift measurements from their optical spectra. Using these distances, one could then calculate absolute luminosities from the observed brightnesses of these objects, which were  used to infer their energy generation mechanisms. From this, 3C273 turned out to be about a trillion times more luminous than the Sun ($\sim\,1\,\times\,10^{12}~L_{\odot}$), and 100 times more luminous than our Milky Way Galaxy.

The radio intensity distribution of the discrete sources were also found to be very different to the optical images of galaxies, nebulae or stars. The very early  interferometers showed that  powerful discrete radio sources (such as Cygnus~A) possessed extended double-lobed structures (\cite{jendas1953}), although many others remained unresolved. Large radio surveys (such as 3C \cite{edge1959}, 4C \cite{pilking1965, gower1967}, MSH \cite{mills1958}) and careful flux density measurements of extragalactic sources indicated that sources with steep radio spectra are associated with extended objects, while compact/unresolved sources preferentially showed flat radio spectra; see Fig~4. of Kellermann (1964) \cite{keller1964}). A complete description of the discrete radio sources required an accurate determination of positional, structural and spectral properties of the sources in emerging radio catalogs. 

There was cosmological relevance to such studies as well. If the extended (double-lobed) radio sources had a preferred linear size (a ``standard rod'') or a fixed rest-frame luminosity (a ``standard candle''), then measuring their angular sizes  and apparent brightnesses over a wide range of redshifts would provide a direct estimation of the curvature of the universe. In addition, the energy generation mechanisms and evolutionary changes over cosmological times could provide  evidence supporting the Big Bang cosmology, (almost established by then, after the discovery of the Cosmic Microwave Background radiation in 1964 \cite{penz1965}), as opposed to the then-attractive Steady State theory \cite{bondi48}. Commissioned into this environment as the world’s largest radio antenna, Arecibo was immediately put to use for projects related to the aforementioned studies. 

With its high sensitivity and (then) finest angular resolution for a single dish, Arecibo's fixed spherical, 305-m diameter primary reflector (see Figure \ref{fig:ao_1963}) was hung within a natural sink-hole.  Its ``chicken-wire'' mesh surface was capable 
of reflecting and transmitting radio waves at decametric wavelengths. Via a steel-cable network, three
 \begin{wrapfigure}{l}{0.275\textwidth}
  \begin{center}
    \includegraphics[width=0.27\textwidth]{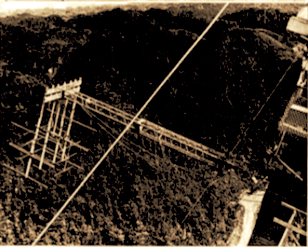}
  \end{center}
  \caption{Feed at the end of an extendable boom, allowing the observation of the Orion Nebula at $-2^{\circ}$ declination}
  \label{fig:ao_boom}
\end{wrapfigure}
$\sim$100-m high concrete towers external to the dish supported a 900-short-ton
triangular feed platform at an altitude of 150~m above the dish. This platform held a symmetric 
bow-shaped ``azimuth'' arm suspended from a huge horizontal circular ring upon which it could rotate in azimuth through 360 degrees. Two so-called carriage houses were hung from a rail track on the azimuth arm on either side of the zenith, allowing them to move in zenith angle. Aberration-correcting line feeds were attached below the carriage houses, which contained receiving and transmitting
equipment. Using synchronized computer-controlled motions of the azimuth arm and the carriage houses, any sky location over the declination range of $-1$ to $+37$ degrees could be tracked for up to $\sim$2.5 h (depending upon the declination).\\
The telescope was equipped with three narrow-band, aberration-correcting line-feeds and receivers 
(any of which could be easily selected) at 318, 430 and 611\,MHz. Point feeds and receivers 
were also available at seven other frequencies. A 10-beam multi-feed receiver system at 611 MHz that was used in large surveys became available later. The paper-tape data recorder used at first light was replaced by a CDC-3300 computer allowing digital data recording from a total power or Dickey-switched radiometer back-end at, for then, a high rate of about 7.4 Mbps. More detail of the telescope and its successive technical upgrades over the years were given by Altschuler and Salter (2013) \cite{altsal2013}. The enormous sensitivity and the possibility of high-speed data recording, shared with the radar work, led to further diverse uses \cite{petten1971}. A few radio astronomy highlights from its first decade are:\\

{\it {\bf Lunar Occultations:}} This technique  was devised by Hazard at Jodrell Bank Telescope and further developed at the Parkes radio telescope (\cite{hazard1961, hazard1962}). In its simplest form,
it can be described as follows: when a small-angular size radio source is occulted by the Moon, the moment of disappearance identifies its position to be somewhere along the  preceding limb of the Moon. Subsequent reappearance, or other occultation events of the same source, give additional such tracks, all of which would intersect at a single point, the most likely position of the source. The locations of the points are limited by timing accuracy and the accuracy to which the position of the limb of the Moon is known, and not by the size of the telescope beam. The total power recordings during these events are actually the straight-edge diffraction patterns of radio sources as they are occulted by the Moon. Inverse transformation of the records  makes it possible to obtain  source-size and structure estimations in a more rigorous manner \cite{scheuer1962}.  The precision achieved by such a method depends on the ability to track the Moon accurately and the signal-to-noise ratio of the recording. At Jodrell Bank and Parkes, only sources stronger than 7 Jy could be observed by this method. Using Arecibo, Hazard aimed to get better positions for sources in the 4C catalog, where the limiting flux density is 2~Jy (at 178 MHz).  The Position errors in the 4C catalog were $1 - 4~s$ in right ascension (RA), and up to $10^{'}$ in declination (Dec). For optical identifications, an error box of $\sim 1 min^{2}$ of arc was required.  Arecibo occultation observations were published in four catalogs \cite{hazard1967,hazard1968, gulkis1969, lang1970a} and with much improved declination values which facilitated optical identifications with distant (extragalactic) objects \cite{hazard1970}. \\

{\it {\bf Interplanetary Scintillations:}} When radio waves from a distant source pass through the electron-density irregularities in our interplanetary solar-wind plasma, intensity fluctuations result at the observer's location due to the relative motion between the scattering medium and the telescope. The angular size of the source modifies the observed  power spectrum of the
intensity fluctuations. By studying such power spectra, (and with reasonable assumptions for the properties of the scattering medium), estimates of source diameters can be obtained at much finer scale than the beam-size of the telescope. The accuracy of such derivations depends only on the signal-to-noise ratio of the observed scintillation power spectra. At Arecibo, a series of such observations for hundreds of discrete radio sources were undertaken mostly by {\it Cohen et al.} \cite{cohen1966a,cohen1966b,cohen1966c,cohen1967a,cohen1967b,cohen1967c,harris1968,cohen1968,cohen1969}.
These helped determine source sizes to fractions of an arcsec. Some of these measurements were also used for solar-wind studies.\\

{\it {\bf Planetary Nebulae and HII Regions:}} Planetary Nebulae were among other important objects studied  during this time at Arecibo.  Measuring their thermal radio emission at low frequencies allowed the  derivation of their electron temperatures using the  optically-thick part of the spectra of detected sources \cite{terzian1966} (for example, NGC~6572, NGC~6720 \& NGC~6853). The unique HII region, the Orion Nebula at Declination $-2^{\circ}$, was also observed by using a special purpose boom that extended out from the carriage house to hold the feed beyond the zenith angle limit of the azimuth arm; see Figure \ref{fig:ao_boom}, \cite{terzian1970}.\\

{\it {\bf VLBI:}} Arecibo also participated in the very early days of Very Long Baseline interferometric (VLBI) experiments (see Section VI).\\

\section{The First Pulsar Epoch: 1968 - 74}
After the paradigm-shifting discovery of pulsars at Cambridge by Jocelyn Bell in 1968 \cite{bell1968}, a flurry of pulsar-related projects began world wide.
Arecibo, with its huge collecting area and fast data-taking systems employed for its planetary radar work, was almost all set for pulsar observations. Consequently, many staff- and visiting-astronomers turned 
\begin{wrapfigure}{l}{0.275\textwidth}
  \begin{center}
    \includegraphics[width=0.265\textwidth]{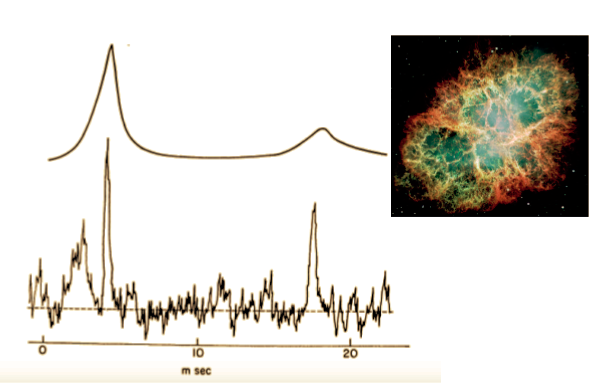}
  \end{center}
  \caption{Radio ({\bf lower}) and Optical ({\bf upper}) pulses from the Crab pulsar from Cocke, Disney, \& Taylor \cite{cdt1969crab_opt}}
  \label{fig:ao_crab}
\end{wrapfigure}
their attention to some aspects of pulsar research \cite{petten1971}. \\
Although the first four Cambridge pulsars were discovered via visual inspection of total-power time series on a chart recorder, Arecibo efforts relied on the digital processing techniques of Fast Folding at a range of trial periods, and on implementing the then-newly-developed Fast Fourier Transform algorithm by Cooley \& Tukey \cite{cooltuk1965}. The first Arecibo  Pulsar was reported by  Craft, Lovelace, \& Sutton \cite{craft1968IAUC} soon after the Cambridge discovery and others followed thereafter. 
Extensive (sometimes first ever) studies of pulse shapes \cite{drake1968, craft1968, rankin1970, backer1970b,couns1971}, dispersion measures \cite{davids1969a, davids1969b, craft1970}, polarization \cite{campbell1970, heiles1970}, pulse ``nulling'' \cite{backer1970b}, intensity fluctuations due to scintillation effects \cite{lang1969, lang1970b, lang1971}, pulse broadening at lower frequencies \cite{rankin1970}, and the monitoring of pulse arrival times \cite{richards1970, rankin1971a, rankin1971b}  were vigorously pursued by several pioneering scientists. \\
An outstanding result came in November 1968 when the period of the pulsating object in the Crab Nebula, NP~0532, was determined to be the then-shortest period of 33.09114 $\pm$ 0.00001 millisecond \cite{lovelace1968}. An average pulse shape obtained at 195 MHz from data corresponding to 25,000 pulse periods gave a pulse width of 3 ms at half intensity and showed a strong post-event feature of about 25\% of the main-pulse intensity, trailing it by 14 ms. This established that a 1.4 $M_{\odot}$ star, rotating 303 times per sec had to have a size of less than just 20 km and hence could only be a neutron star, and not a white dwarf \cite{gold1968Nstar}. They also established, that out of the two compact (variable) objects in the vicinity of the Crab Nebula, discovered earlier at NRAO Green Bank, NP~0532 was at the right distance (estimated from its dispersion measure) to be associated with the nebula, and hence  directly related to the Supernova of CE. 1054. The accurate pulse period measurement was an essential factor in the heroic optical identification of Baade's star with the pulsar and the detection of its optical pulsations having the same period by Cocke, Disney, \& Taylor \cite{cdt1969crab_opt} (Fig. \ref{fig:ao_crab}). These results completed the story of high-mass stellar evolution leading to a violent supernova explosion leaving behind a rotating neutron star, i.e. pulsar, at the center of a remnant \cite{drake1971}.\\
The next big step in pulsar search methods was the development of the FFT search with Incoherent Harmonic Summing by Joe Taylor, who employed this at Arecibo along with his graduate student, Russell Hulse. This method formed the basis of almost all following pulsar search programs. As a result, 150 new pulsars were found in the first eight years to 1976 using radio telescopes at Molonglo \cite{vaughan1969,large1971,vaughan1972}, Jodrell Bank \cite{lyne1968Nat,davies1970a,davies1970b}, and Arecibo \cite{hulse1974}.\\
The ``Jewel in the Crown'', however, was the first binary pulsar PSR~B1913+16 discovered at Arecibo in 1974 by Russell Hulse and Joe Taylor \cite{taylor1974IAUC}. The observations actually took place during the Arecibo primary surface upgrade (Sec. III). The two neutron stars in the system have almost equal masses of 1.4 $M_{\odot}$, of which only one is seen as a pulsar from the Earth. They follow elliptical orbits around their center of mass with a 7.75 hr periodicity.
\begin{figure}
 \includegraphics[width=0.35\textwidth]{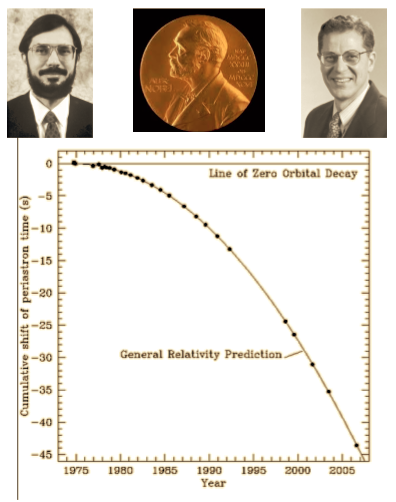}
 \caption{Top row: photos commemorating the 1993 Nobel Prize in Physics. Bottom panel: binary pulsar B1913+16, with shrinking orbit following general relativity. From Weisberg et al., (2010) \cite{WNT2010}.}
\label{fig:binpsr}
\end{figure}
Observations of PSR~B1913+16 over the next few years  established that the binary orbit was shrinking by $\sim1~cm/day$, with the system losing energy due to gravitational radiation as per Einstein's General Theory of Relativity. This important result led to the award of the 1993 Nobel Prize in Physics to Hulse and Taylor (Figure \ref{fig:binpsr}).
\section{Arecibo joins HI \& OH Research (1974 - 92)}
Towards the end of Arecibo's first decade, it was decided that the 96-m, 430-MHz line feed would be replaced by one of greater efficiency. The existing ``chicken-wire mesh'' surface was also to be replaced by perforated aluminum plates, allowing operations at much higher frequencies than previously. 
Thus started the process of Arecibo reinventing itself a number of times  over the next 40-plus years of its life, keeping itself at the forefront of research.
\begin{figure*}
   \centering
    \begin{minipage}{0.47\textwidth}
        \centering
        \includegraphics[width=0.95\textwidth]{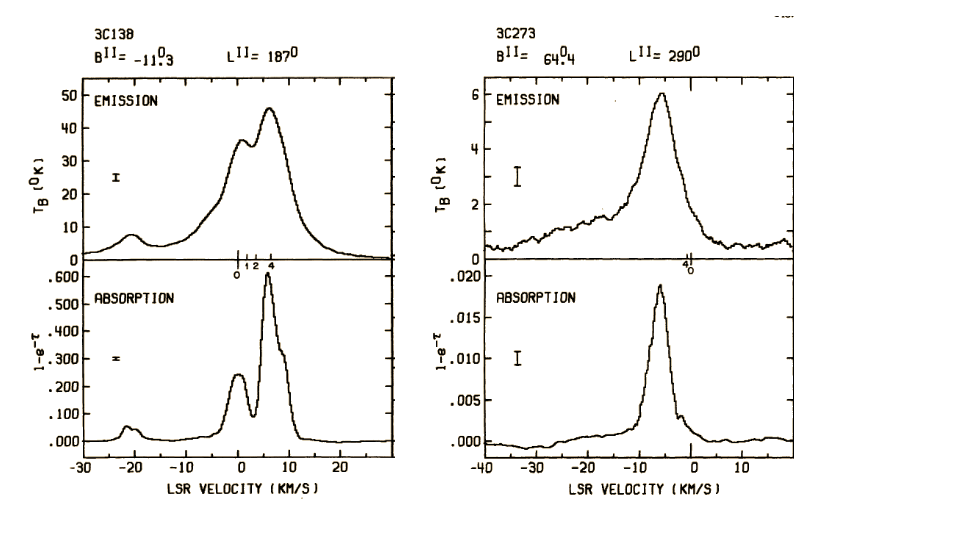} 
        \caption{HI emission-absorption spectra toward 3C138 and 3C279. Lines of sight to these extragalactic radio sources slice through very different Galactic mixing of WNM and CNM clouds.
        From Dickey, Salpeter, and Terzian, 1978 \cite{dickey1978}}
        \label{fig:hiem_abs}
    \end{minipage}\hfill
    \begin{minipage}{0.51\textwidth}
        \centering
        \includegraphics[width=0.95\textwidth]{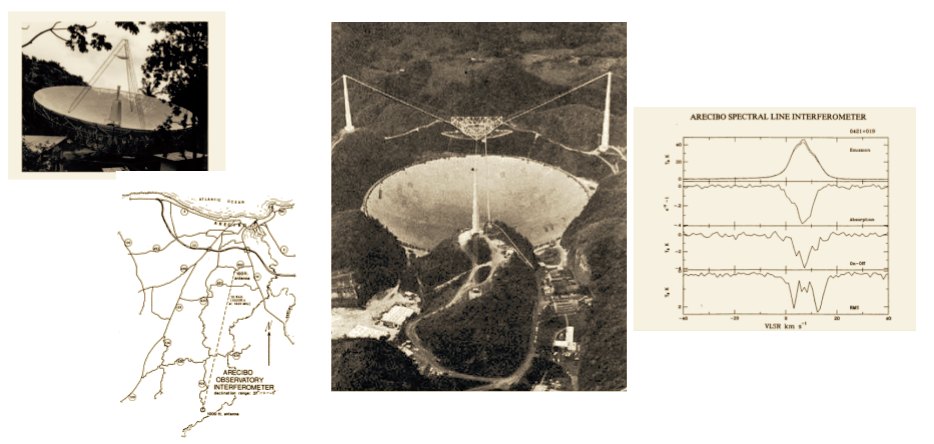} 
        \caption{The Arecibo-Los Ca\~nos interferometer and an example of a spectrum taken with it.
        From Kulkarni et al. (1985) \cite{Kulk85}.}
        \label{fig:los_caneos}
    \end{minipage}
    
\end{figure*}
Radio astronomy flourished at Arecibo over the next two decades. The hyperfine transition of the neutral hydrogen atom (HI) at 1420.4058 MHz was accessible to the large antenna for the first time. It was possible to observe redshifted HI to a recessional velocity of $\sim 22,800~km s^{-1}$, with a ``22-cm feed'' covering the frequency range 1310 - 1390 MHz. The 18-cm OH radical ground-state transitions were also covered. An auto-correlation  spectrometer was provided as the datataking back-end. Among the major areas of research  at that time, areas that were explored throughout the rest of the telescope's life were studies of the interstellar media (ISMs) of both our Galaxy and external galaxies.\\

\textbf{\emph{HI in the Galactic ISM:}} When Arecibo entered  the field of ISM research via the 21-cm line, there had already been a plethora of Galactic HI studies over the previous 15 years, mostly using single-dish radio telescopes with beams of around 1$^{\circ}$; see reviews by Burton (1974)\cite{burt1974} and Verschuur (1974) \cite{vers1974} and references therein. Face-on maps of the outer Galaxy had  been constructed mostly from Dutch and Australian observations using the Doppler shift of the HI emission lines in combination with the Galactic rotation curve (e.g. \cite{vdh1954, wester1957, oort1958}.) Such studies revealed the concentration of the HI gas in spiral arm features, although the detailed  density and temperature structures were unknown owing to the complexity of the velocity field, and the overlapping of multiple components \cite{kerr1969}. \\

The discovery of HI absorption in the spectra of discrete sources (\cite{hlm1955})
indicated the presence of cold absorbing gas ``clouds'' along a line-of-sight through the Galaxy. 
In the presence of other known constituents such as the ionized medium producing the dispersion and scattering of pulsar signals, supernova remnants, HII regions around newly born stars, and large out-of-plane shell-like structures (e.g. \cite{heiles1979}), a comprehensive description of the physical state of the ISM was required.\\ 
Two major theoretical models involving multiple phases were advanced that attempted to achieve a stable equilibrium considering the heating and cooling of the ISM. The first, by Field, Goldsmith, \& Habing \cite{fgh1969} had a cold, dense cloud phase [Cold Neutral Medium (CNM), with $T < 300K$], surrounded by a warm intercloud phase [the Warm Neutral Medium (WNM), with $ T \sim 10^{4} K$] of neutral and ionized gas. Later, McKee \& Ostriker \cite{MO1977} included a third, dynamic phase of hot gas, heated by shock waves from supernovae [the Hot Ionized Medium (HIM), with $T \sim 10^{6} K$], filling the majority of the 
ISM.\\
The observing method for studying the WNM and CNM was either emission-absorption spectroscopy towards extragalactic radio sources, or self-absorption in the WNM clouds (\cite{rad1972}). The limited angular resolution  used until then prevented the exact measurements of the temperatures in the CNM as the emission was extended and beam filling, while the HI-absorbing clouds were mostly much smaller in angular size, and multiple clouds with different velocity ranges could also be included within the beam. The need for finer spatial and velocity resolution was apparent.
At Arecibo a series of emission-absorption spectra were taken with 3.3$'$ angular resolution and 0.1 $km\,s^{-1}$ velocity resolutions \cite{dickey1978, crovis1980}; see Fig. \ref{fig:hiem_abs}). To further improve the spatial resolution,  a two-element, radio-linked interferometer was specially constructed with a baseline of about 10 km between the 305-m main dish and an equatorial-mounted 30-m parabolic dish at Los Ca\~nos; see Fig. \ref{fig:los_caneos}). This resulted in an angular resolution of $4^{''}$, and removed the effect of large-scale HI variations contributing to single-dish ON-OFF measurements when obtaining the emission-absorption spectra of compact background radio sources. It helped sampling
almost the same 3-D volume (comprising two spatial dimensions and a velocity dimension) of the ISM for measuring the emission-absorption spectra \cite{Kulk85}.\\
These Arecibo measurements provided the most reliable values for the temperatures and column densities of the CNM clouds \cite{crovis1981} up until the late 1980s, although their sizes were not well constrained (they were dependent on optical depth estimates). The very first measurement of AU-sized absorbing clouds came from VLBI absorption observation toward 3C147 (\cite{diet1976}) using a Hat Creek to Owen's Valley baseline providing  0.1 $^{'}$ angular resolution. Fourteen years later, these observations were corroborated by Diamond et. al. \cite{diam1989}, while HI spectral-line observations at Arecibo toward pulsars \cite{frail1994} indicated similar  structures. For the WNM, owing to the ubiquitous nature of the HI-emission, and the huge shell and super-shell structures seen in early maps, synthesis imaging was not of great use in studying these higly extended objects. 
Much additional information on the CNM, WNM, and the interstellar magnetic field came from the Millennium Arecibo 21-Centimeter Absorption-Line Survey \cite{heiles2003}. The understanding of the overall structure of the HI in the ISM changed dramatically as a filamentary structure was revealed by the GALFA-HI survey using the seven-beam Arecibo L-band Feed Array (ALFA) receiver on the Arecibo telescope after the Gregorian upgrade. We return to that story in Sec. V.\\

\textbf{\emph{Extragalactic HI, the large-scale distribution of galaxies,  \& the Perseus-Pisces Supercluster:}}
A few important results drawn from observations made at different radio telescopes in the field of  21-cm HI studies of galaxies spurred the initial projects at Arecibo as it entered the field after the first upgrade. These are, (1) studies of the HI properties of galaxies along the Hubble sequence \cite{roberts1969}, (2) the flat rotation curves of spiral galaxies implying the existence of ``Dark Matter'' \cite{rubin1978} and (3) redshift-independent distance estimations of galaxies using the Tully- Fisher relation. We now describe the contributions made by Arecibo to these and related fields of HI in external galaxies.\\
Using optical and 21-cm radio spectral-line observations, (taken by the Harvard, Leiden, Nancay, Parkes, and Green Bank groups), Roberts \cite{roberts1969} explored how the median values of $M_{HI}/L_{ph}$, $M_{HI}/M_{total}$, and $M_{total}/L_{ph}$ and surface densities of $L_{ph}$, $M_{total}$ and $M_{HI}$ (where, $M_{HI}$ \& $M_{total}$ are a galaxy's HI and total mass and $L_{ph}$ is its photometric luminosity) vary with morphological type from Sa to Ir. However, their sample size was just 91 galaxies for which all the required data were then available. The early-type (S0, \& E) galaxies were mostly undetected. It seemed important, given the huge sensitivity boost (about a factor of 10 over the 300-ft telescope at Green Bank) that the upgraded 305-m telescope be applied to extend this study. \\
\begin{figure}
 \includegraphics[width=0.45\textwidth]{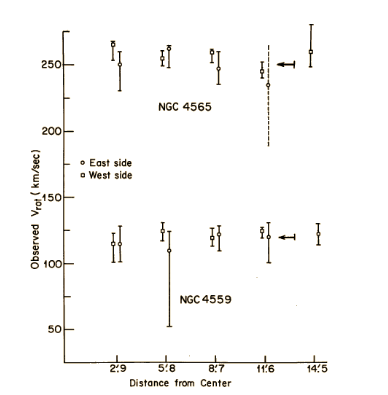}
 \caption{Galaxy rotations curves out to radii of $\sim 80~kpc$. From \cite{krumm1977}}
 \label{fig:ao_rot_curv}
\end{figure}

Three major groups pursued this area of investigation. Among them, Krumm \& Salpeter \cite{krumm1976} observed early-type galaxies, including the E, S0 and S0/a galaxies. Bieging \& Biermann (\cite{bieging1977}) observed a large sample of S0 galaxies, while a very sensitive  search for HI emission from 38 mostly ellipticals was undertaken by Knapp et al. \cite{knapp1978}.  For the first time,  the ratio between the HI mass and B luminosity, M(HI)/L(B), of S0 galaxies was derived and found to be less than 0.1 for all ``true'' S0 galaxies. This and the colors observed for the S0 galaxies were consistent with the suggestion that weak bursts of star formation occur in these galaxies and may account completely for their observed low HI contents. In contrast, elliptical galaxies were found to have very little HI.\\
The flat rotation curves of spiral galaxies and the implied existence of ``Dark Matter'' was first presented at the 1975 AAS meeting by Dr. Vera Rubin (with subsequent publications in 1978, \& 1980 \cite{rubin1978, rubin1980}). Stars far away from  the Galactic bulge were observed to have the same rotational velocities as their counterparts closer to  the central regions, implying that the  gravitational mass of galaxies grew linearly with galacto-centric distances. If Newtonian gravity were to apply to such systems, then 50\% or more of the mass must exist in the form of dark halos. At Arecibo, Krumm \& Salpeter \cite{krumm1977} observed 14 edge-on spirals of type Sb to Sm with $R_{opt} > 3'$, and blue magnitudes brighter than 11. In six of these, HI was detected out to more than $10'$ or 80~kpc from the center of each galaxy. Flat rotation curves were seen in all of these (Figure \ref{fig:ao_rot_curv}), strengthening the conclusion of Rubin et al.\\
\begin{figure*}
 \includegraphics[width=0.9\textwidth]{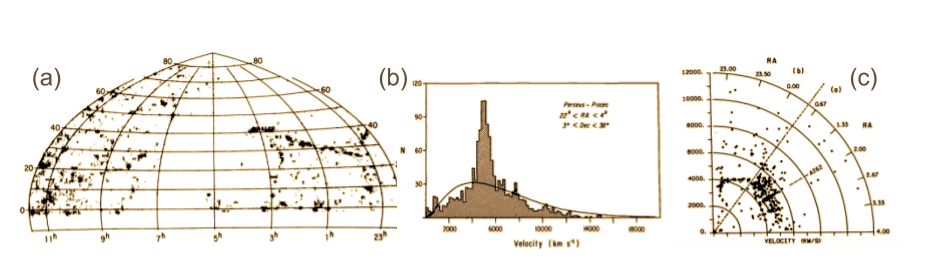}
 \caption{Galaxy distribution in the Perseus Pieces supercluster: {\bf (a)} projected distribution from the CGCG optical catalog, {\bf (b)} 21-cm redshift histogram from \cite{gio1983IAUS}, {\bf (c)} Pie distribution with a delta-dec. of $4^{\circ}$, showing the velocity range of 3000 - 6000 $km\,s^{-1}$ confining the filament in the third dimension.From Giovanelli, 1983 \cite{gio1983IAUS}.}
 \label{fig:gal-3d-pp}
\end{figure*}
The Tully-Fisher  relation \cite{TF1977} represents an empirical inter-dependency between two distance-independent quantities, the 21-cm line width and the absolute luminosity of galaxies. Using this, together with the observed optical apparent magnitudes, the distances to  galaxies could be determined and compared with the values from the (redshifted) frequency of their HI lines. This, in turn, provided an independent estimate of the Hubble constant ($H_{\circ}$), and helped in studying the large-scale 3-D distribution of galaxies beyond the local region.\\
An extensive study by Giovanelli, Haynes and collaborators was summarized in a 1983 publication \cite{gio1983IAUS} where 1100 galaxy-redshift measurements were made at Arecibo. They constructed a 3-D distribution of the long, filamentary, over-dense region seen in projection of the galaxies of the Catalogue of Galaxies and of Clusters of Galaxies (CGCG) by Zwicky, Herzog, and Wild (1968)\cite{cgcg1968}. The data of Giovanelli, Haynes and collaborators were also used to consider the environmental effect on the derived value of the Hubble constant in the original Tully-Fisher study, in which they used most of the published HI spectra up to that time and their own observations using the NRAO 91-m and 140-ft telescopes at Green Bank (see Fig~1A, and the Appendix of \cite{TF1977}). By 1992, the number of galaxies in this Arecibo Perseus-Pieces catalog had grown to contain about 8250 21-cm redshift entries (out of 30000 total measured redshift values for galaxies using both radio and optical methods), which made its place secure among the preautomated, fiber-optic, multiobject spectroscopic redshift surveys. This study helped to establish the large-scale distribution of galaxies forming filamentary overdense regions and ``voids'' of galaxies (Figure \ref{fig:gal-3d-pp}). However, untargeted (``blind'') HI surveys for galaxies had to wait for yet another technical advance at Arecibo, which made a variety of galaxy-related studies possible, and which is still ongoing using archived ALFA data (see Sec. V.B).\\

\textbf{\emph{Envelopes of AGB  stars via OH satellite lines:}} Toward the end of their evolution, intermediate-mass stars go through a stage of high mass loss. These asymptotic giant branch (AGB) stars then form a thick, cold  circumstellar envelope of gas and dust containing several molecular species. The envelope reprocesses the stellar radiation into the infrared and under suitable circumstances produces maser emission. Until the early 1980s, these stars were discovered and studied via their 1612-MHz OH satellite lines and then found to have strong IR emission. Results from the space-based Infrared Astronomical Satellite (IRAS) \cite{nauge1984a} changed this situation dramatically. Using the IRAS Point Source Catalog, it was realized that the then-known OH/IR stars occupied a definite region in the (60~$-$~25) $\mu$m versus (25~$-$~12) $\mu$m color-color diagram (\cite{bml1985Nat}). Lewis et al.  observed the 1612-MHz OH line for a complete sample of such IR objects within the Arecibo sky. In three subsequent publications \cite{ohir11988, ohir21990,ohir31993}, they presented the results for the observations of 1900 sources with 213 new detections, thereby doubling the number of known OH/IR stars by 1993.  Many of these were later monitored to follow the temporal evolution of their shells. This has provided important input for constructing a generalized chronological sequence of appearance of $SiO$, water, OH-mainline, and 1612-MHz satellite line masers \cite{bml1989} that describes the ultimate formation of proto-planetary nebulae from these stars.\\

\begin{figure}
 \includegraphics[width=0.5\textwidth]{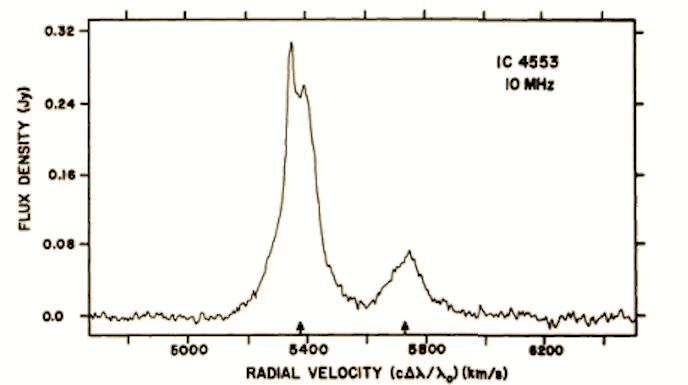}
 \caption{First-ever detection of an OH megamaser in the ULIRG Arp220. From Baan, Wood, and Haschick, 1982 \cite{baan1982}.}
 \label{fig:ohm1}
\end{figure}

\textbf{\emph{The First OH megamasers in external galaxies:}} The first-ever extragalactic OH maser emission, a million times stronger than those in  Galactic-star-forming regions such as W3(OH), was detected at Arecibo in the peculiar galaxy Arp220 by Baan et al. \cite{baan1982}. This marked a new subclass of radio sources, OH megamasers (OHMs). This discovery was made somewhat serendipitously (Figure \ref{fig:ohm1}) during a search for OH absorption in galaxies with strong HI absorption \cite{mirabel1982}. Later, IRAS data established a class of objects named ultraluminous infrared galaxies (ULIRGs) having $L_{FIR} > 10^{12} L_{\odot}$ \cite{soifer1984}, of which Arp220 became the prototype. In most of the ULIRGs, the far-infrared (FIR) flux appears to have come from dust that was heated as a consequence of an intense burst of star formation. However, at the highest range of FIR  luminosities, this class of objects could be powered by both starbursts and active galactic nuclei (AGNs), and has been found in interacting or merging galaxies \cite{elston1985,lawrence1988}.\\
Further searches for OHM emission after the Gregorian upgrade in 1997 (Sec. V) resulted in about 120 OHMs among the ULIRG population \cite{darl2002, klock2004, zhang2014}. OHMs could be observed out to a much larger distance ($ 0.1~<z<~0.45$) at Arecibo compared to the HI emission in normal galaxies, although only about 25\% of ULIRGs are OHM sources. An evolutionary scenario originally put forward by Sanders et al. \cite{sanders1988} is currently favored. It links ULIRGs to other systems, whereby a merger of two gas-rich spirals forms a luminous infrared galaxy that evolves into a ULIRG, then a quasar, and eventually an elliptical galaxy. \\
\section{Millisecond Pulsars \& Exoplanets (1982 - 92)}
Two major discoveries bracketing the 1980s changed the course of research trajectories, with monumental impacts in  their sub-fields. These are the discoveries of millisecond pulsars \cite{backer1982,backer1982Nat},  and exoplanets \cite{wolsz1992Nat}.\\
\begin{figure*}
 \includegraphics[width=0.85\textwidth]{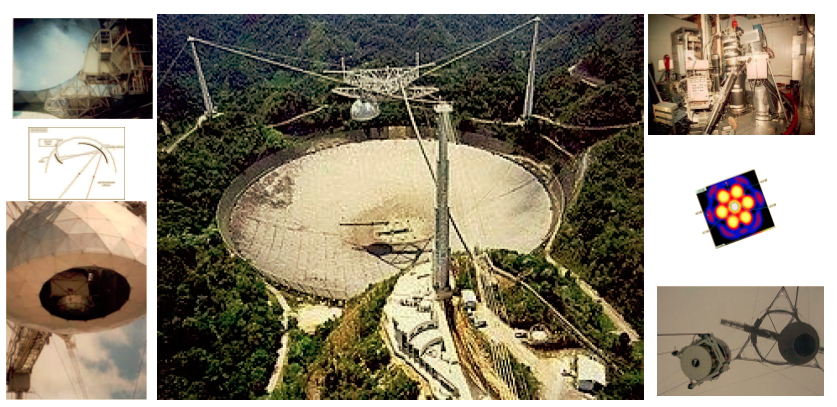}
 \caption{Left images: the Gregorian optics. Center image: the upgraded telescope in 1997. Right images: the seven-beam ALFA receiver and its beam pattern on the sky. ALFA was installed in 2005. Courtesy of Arecibo Observatory}
 \label{fig:upgrade2}
\end{figure*}
\textbf{\emph {The first millisecond pulsar}}  was found at Arecibo in 1982 by Backer et al. due to their sheer persistence in trying to solve an astronomical mystery that had existed for over five years. Backer recounted the story soon thereafter \cite{backer1984}, detailing how 1.533-millisecond pulsed emission was finally detected from the highly unusual, compact, steep-spectrum, strongly polarized radio source 4C21.53W. The key was to look beyond the parameter space occupied by all then-known pulsars (somewhat defying the contemporary theories of pulsars) and sample the data at a rate  of 0.5 ms. Backer et al. observed at 1400 MHz, where scattering due to the interstellar ionized medium would have less effect than at 430~MHz, where most previous Arecibo pulsar work had been carried out.\\
A second millisecond pulsar PSR 1953+29, with a period of 6.1 ms was discovered at Arecibo shortly afterward by Boriakoff et al. \cite{boria1983IAUC}. The spin-down rate of these pulsars were exceedingly small, $<~10^{-19} ss^{-1}$, and had extremely stable periods and low magnetic fields. They were later shown to be a new class of old (``recycled'') pulsars that are spunup in binary systems \cite{rad1982}. The total number of known millisecond pulsars as of May~2024 was 552, about 20\% of which were discovered at Arecibo \footnote{{ https://www.astro.umd.edu/$\sim$eferrara\\/GalacticMSPs.html}}.
In 1990, Foster \& Backer \cite{fosbec1990} pointed out that since millisecond pulsars are the most accurate time-keepers,  analysis of their pulse arrival times could be used (i) to provide a time standard for long time scales, (ii) to detect perturbations of the earth's orbit, and (iii) to search for a cosmic background of gravitational radiation.  They described how  these  effects could  be estimated and applied to the  analysis of timing data. They also started a nascent ``pulsar timing array'' that turned out to be  instrumental in the detection of the stochastic background gravitational waves via  projects such as North American Nanohertz Observatory for Gravitational Waves (NANOGrav); see Sec. VIII. 
Demorest and Goss (2024)\cite{demorest2024} recently published an excellent review of the discovery of millisecond pulsars providing a ringside view of not only this discovery but also the varied applications of these objects in a broad range of fundamental physics questions.\\
\textbf{\emph {The ``Black Widow'' pulsar:}} In 1988, the first of a notable class
of millisecond pulsars, PSR B1957+20 was discovered at Arecibo (Fruchter, Stinebring, and Taylor, 1988 \cite{fruchter1988}). Termed the ``Black Widow'' pulsar, the companion in this eclisping binary 
system, a brown dwarf or a super-Jupiter, is in the process of being  destroyed by the strong outflows of high-energy particles  from the neutron star. This pulsar provided a wealth of information on the evolutionary history of millisecond pulsars. By early  2023,  around 41 black widows were known to exist.\\

\textbf{\emph{The first  exosolar planets:}} By 1992, these were also discovered at Arecibo via an accurate pulse-arrival-time analysis of a 6.2 millisecond pulsar PSR B1257+12 ((Wolszczan and Frail, 1992 \cite{wolsz1992Nat}). One moon-sized and two terrestrial-mass planets orbit around this pulsar in  similarly inclined orbital planes, strongly suggesting their common origin in a protoplanetary disk.
This was the first direct proof of the existence of exosolar planets. Currently, several thousand planetary systems are known, with about half a dozen around pulsars.

\section{The Gregorian Upgrade (1992 - 97)} The middle of the 1990s saw yet another major upgrade of the Arecibo telescope and its entire suite of electronics and computing systems. Except for the primary surface of the dish, everything was updated to contemporary technology. A quest for the perfect geometric focus had been on the minds of Arecibo's leaders from day one. By the late 1980s, the success of a prototype ``mini-Gregorian'' project \cite{magnani1998} provided the proof of concept for a dual-reflector feed in Gregorian optics to be able to do just that \cite{kildal1991}. By mid-1996 two shaped reflectors within a large dome, hanging from the azimuth arm, brought the rays reflected off the primary spherical dish to a geometric focus, thereby rendering the telescope achromatic. A 50-ft-high stainless steel ground screen around the dish made the system temperature ($T_{sys}$) far less dependent on the telescope Zenith Angle by deflecting  the vignetted beam onto the sky, where the contribution to $T_{sys}$  is 2.7~K from the cosmic microwave background (except at low frequencies), and not 300~K from the surrounding hills. 
Multiple state-of-the-art receivers covering 0.3 to 10 GHz were located on a rotatable turret in the Gregorian dome to allow frequency agility. Any of these could be brought to the focal point and signals piped down to a user-chosen backend via  reconfigurable electronic circuitry, and eventually via a highly efficient graphical user interface named CIMA. This allowed for remote observing over a secure, high-speed Internet connection to the Observatory.  An on-loan Jet Propulsion Laboratory (JPL) H-maser was deployed at the heart of the observatory's timekeeping and frequency-standard system; up to that point, Arecibo was using a rubidium clock. World-class VLBI data-acquisition and recording systems brought Arecibo's enormous sensitivity to the service of VLBI arrays; see Sec. VII. In 2005, a 7-beam L-band receiver system (developed at the Commonwealth Scientific and Industrial Research
Organisation in Australia) and a matching wideband spectrometer were deployed, opening up the telescope for use by various key surveys (Fig. \ref{fig:upgrade2}). Thus began the “modern era” of radio astronomy at Arecibo. During this time, radio astronomy blossomed there in almost all of its subfields and continued apace until August 2020, when the first auxiliary-cable break suspended operations for good.\\
\subsection{Arecibo at ``High'' frequencies}
After a five-year long hiatus for the Gregorian upgrade, when the Arecibo telescope restarted observing in 1998, a large number of pulsar projects were on the telescope immediately. Four different data processing backends, employing different digital techniques were deployed, some of which were ``user-owned-public-access'' devices, to complement the facility instrument developed at the Observatory. The projects ranged from the confirmation of  candidate pulsars found during the upgrade-time 430-MHz drift-scan survey, through searching for pulsars in globular clusters and objects associated with Fermi gamma-ray pulsars, to the accurate timing of all known (and visible from Arecibo) millisecond pulsars to ensure the continuity of the phase of the pulses over the five-year-long ``lost'' time. This period also saw pulsar proper motion measurements via VLBI techniques and the start of a new 327-MHz drift-scan search. Taking advantage of the frequency agility and the receivers above the S-band, {\bf molecular-line transitions} other than the L-band  OH radical began to be utilized for astronomical studies for the first time at Arecibo. A few examples of these studies follow:\\
\begin{figure*}
\includegraphics[width=0.95\textwidth]{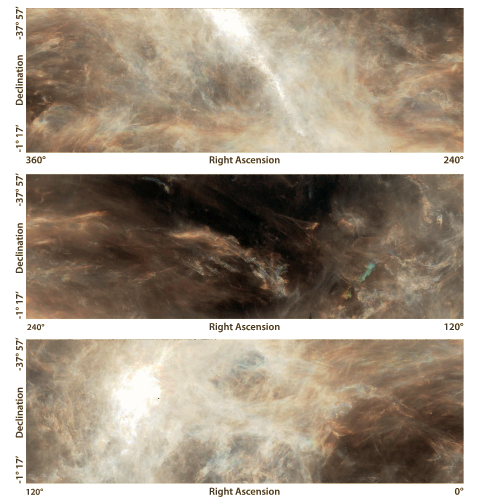}
\caption{Visualization of the GALFA-HI data. Three 1 $km~s^{-1}$ velocity slices are shown at $V_{LSR}$\,=\,+2, +3, and +4 $km\,s^{-1}$ in blue, green, and red, respectively and their brightness temperatures are scaled logarithmically over a range of 1 -- 100 K. From Peek et al., 2018 \cite{peek2018}} 
\label{fig:galfa_HI}
\end{figure*}
\textbf{\emph{Simultaneous radio recombination line (RRL) emission and formaldehyde ($H_{2}CO$) absorption towards Ultra Compact HII regions (UC\,HII) \cite{araya2002}:}}. Newly formed high-mass stars cause ionized regions around themselves that are deeply embedded within the molecular cloud from which the stars have formed. These are small, with diameters $\le 10^{17}cm$ and dense (with number density $>10^{4} cm^{-3}$) and are visible only in the IR, submillimeter and radio regimes.  Given their density, these HII regions should be short lived, yet they are numerous, hence long lived and among the most luminous objects at $ \lambda\,100~\mu\,m$ in the Galaxy (the ``age crisis'', \cite{wood1989}). Using simultaneous measurements of C-band RRLs  and the 4.829-GHz transition of the $H_{2}CO$ molecule, Araya et al. (2002) \cite{araya2002} established the velocities of a sample of these UC\,HII regions. The $H_{2}CO$ absorption spectra  were used to resolve the distance ambiguity for each source. Even with the limited number of UC\,HII regions, they found good agreement with  spiral arm locations. This implied that interactions with the molecular cloud need to be taken into account while deriving the age of such HII regions.\\

\textbf{\emph {High-frequency observations of carbon RRLs:}} A search for the carbon RRLs toward 17 ultracompact HII regions was carried out at 8.5 GHz \cite{roshi2005}. These were the first-ever X-band spectral line  observations at Arecibo. Eleven sources indicating the presence of dense photodissociation regions (PDRs) associated with the UC\,HIIs were detected, providing important information on the kinematics and physical properties of their ambient medium.\\
\begin{figure*}
\includegraphics[width=0.95\textwidth]{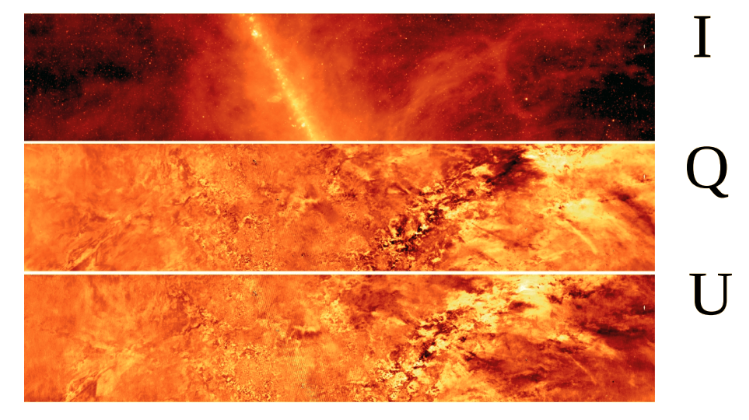}
\caption{\label{fig:galfacts}Preliminary GALFACTS images of the Galactic Center quadrant in three Stokes parameters, I, Q, and U. From  Taylor, A. R., 2015 (private communication).}
\end{figure*}
\textbf{\emph{A 4-6 GHz spectral scan and specific spectral-line observations within 8–10 GHz of the dark molecular cloud, TMC-1:}}
In a survey of the lowest-frequency molecular lines toward the  cyanopolyyne peak of the Taurus Molecular Cloud-1 region, the already-known $H_{2}CO$ and $HC_{5}N$ lines and molecular lines of $C_{2}S, C_{3}S, C_{4}H, C_{4}H_{2}, HC_{3}N, H^{13}C_{3}N$, as well as of $HC_{5}N, HC_{7}N$, and $HC_{9}N$ were found \cite{kalen2004}. About half of these were detected for the first time. These highly sensitive observations helped determine the rotational temperatures of the detected molecules to be in the range 4-9 K and  the Cyanopolyyne column densities to vary from $5.6~\times~10^{13}\,cm^{-2}$ for  $HC_{5}N$ to $2.7~\times~10^{12}\,cm^{-2}$  for $HC_{9}N$.\\

\textbf{\emph{An unbiased methanol maser survey of the Galactic plane:}} This was conducted by Pandian et al. \cite{pandian2007a,pandian2007b} in order to trace high-mass star formation sites that were believed to be the precursors of UC\,HII. The survey covered a region of the Galactic plane in the  longitude range $35^{\circ}\,<\,l\,<\,55^{\circ}$, and latitudes $|\,b\,|~\leq~0.4^{\circ}$ and found 86 sources. With a $3\sigma$ sensitivity of 0.25~Jy, it was at the time (2005) the most sensitive methanol maser untargeted survey. The catalog formed the basis of the luminosity function for Methanol Maser sources in the Galaxy. For this survey Pandian et al. built a low-noise, cooled receiver for the Arecibo telescope  to cover the frequency range of 6 to 8 GHz, filling a then nation-wide gap for all U.S. radio astronomers. \\

\subsection{The ALFA surveys (2005 onward)} 
In 2005 the seven-beam ALFA was installed, and associated large-scale surveys were set to begin. To maximize the efficiency of these surveys, the observatory organized several community meetings where three major consortia were formed, with each focusing on closely related science goals. These were GALFA, PALFA, and EALFA, covering Galactic, pulsar and extragalactic astronomy, respectively \cite{alfasurveys2007}. Each, with membership open worldwide and consisting of $\sim$50 participants per group, developed and executed the observing strategies, data reduction, the derivation of scientific results and their public dissemination. Each of these also had large numbers of students and post-doctoral fellows involved. The data were gathered from 2005 until well into 2013, with follow-up projects continuing until the final days of operation of the telescope. Some of the data accumulated by the consortia are still being reduced and analyzed. We now highlight a few general results from these surveys.\\

\textbf{\emph{The Galactic ALFA HI (GALFA-HI) survey:}} This survey covers the entire Arecibo sky from declination $-1^{\circ}~17^{'}$ to $+37^{\circ}~57^{'}$ across all RA (13,146 sq. deg.), recording HI spectra within the velocity range of $-650$ to $+650~km\,s^{-1}$,
with  0.184 $km\,s^{-1}$ velocity resolution. It had $4'$ angular resolution, and 150 mK rms noise per 1 $km\,s^{-1}$ velocity channel \cite{peek2018}. Half of this survey was conducted commensally with the
GALFA Continuum Transit Survey (GALFACTS); see the subsequent discussion. Figure \ref{fig:galfa_HI} shows a limited view of the completely reduced data demonstrating the filamentary nature of the general composition of the Galactic HI. The data are publicly available online \footnote{{ http://stsci.edu/$\sim$jegpeek/galfadr2}} and through the Harvard Dataverse \footnote{{see 10.7910/DVN/T9CFT8}}. The original consortium listed several areas of scientific inquiries that intended to use this data. \\
In contrast to the earlier three-phase model of the ISM (see Sec. III), this survey shows that the HI medium is dominated by filamentary structures. Further analyses \cite{vers2018a, vers2018b}  indicated that the HI features cannot be in a simple thermal pressure balance,  and appear to
require a magnetic field component, with a strength of a few to a few tens of $ \mu$G, to confine the filaments and explain the observed substructure. Efforts are ongoing to detail the process via which the neutral HI interacts with the Galactic magnetic field (\cite{clark2014, campbell2022, kim2023}). \\

\textbf{\emph{The GALFA Continuum Transit Survey, GALFACTS:}} This is a sensitive, high-angular  resolution, spectropolarimetric continuum survey, imaging the entire Arecibo visible sky \cite{galfacts2010}. Beginning in early 2000, highly structured, arcminute-scale, polarized radiation had been seen in interferometric images of the nonthermal continuum background, both on and off the Galactic plane (\cite{haver2003, taylor2003, land2010}). These structures had no counterparts in Stokes~I, the total intensity. GALFACTS presented an opportunity to image the continuum radiation in full polarization with a Single Dish, which safeguarded against the missing short (to zero) spacing information of interferometric maps, thereby retaining  the large-scale distributed structures. Among additional goals were to image the polarized emission from discrete objects and  
the diffuse interstellar medium of our Galaxy, and to obtain Faraday rotation measures for a large number of extragalactic sources. This explores the role that the magnetic field plays in the diffuse structure of the Galaxy (Fig. \ref{fig:galfacts}).\\

\begin{figure}
 \includegraphics[width=0.45\textwidth]{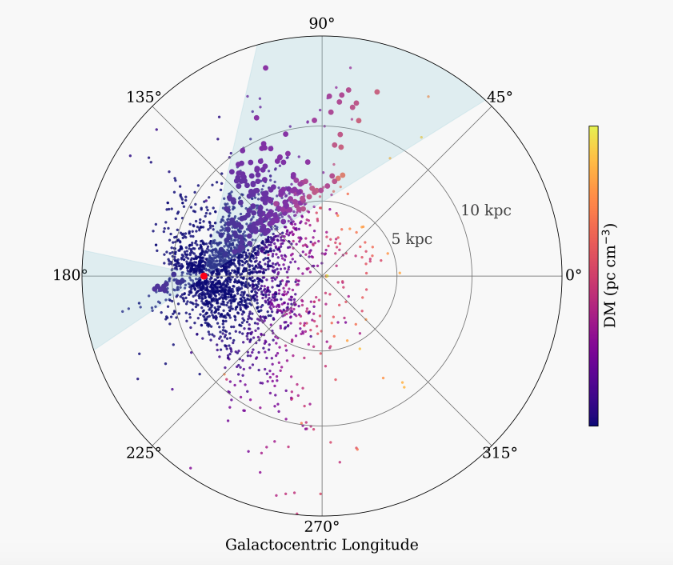}
 \caption{\label{fig:palfa_dist} The Galactic location of new pulsars discovered in the PALFA survey marked by larger dots.  The position of the Solar System is indicated by the red dot, and the search areas are indicated in light blue. The dots indicate all known pulsars, colored according to their 
 ispersion measures (DMs). From Parent, 2025, \cite{parent2019} and Wikipedia, 2025.}
\end{figure}

\textbf{\emph{PALFA:}} PALFA has been a large-scale survey to discover pulsars and transients with the Arecibo telescope  covering the reachable sky within $5^{\circ}$ of the Galactic plane \cite{cordes2006}. The data acquisition initially had a 0.1-GHz, and then, after 2008, a 0.3-GHz, bandwidth at 1.4 GHz. With time sampling of  $\sim 64\,\mu sec$ and dispersion smearing limited to 1.2 ms, the survey was well suited for finding millisecond pulsars. As of 2019, the survey had yielded 208 new pulsars (Fig. \ref{fig:palfa_dist}), including three double neutron star systems \cite{lori2006, vanL2015, laza2016, stov2018} and 35 highly recycled millisecond pulsars ( \cite{champ2008,  crawf2012, deneva2012, knis2015, schol2015, stov2016}). Of these, PSR J1903+0327 and PSR J0557+1551 have been included in the NANOGrav Pulsar Timing Array (\cite{arzo2018}). Data from these surveys are being actively used for various follow-up studies \cite{parent2019}.\\

\textbf{\emph{Extragalactic HI surveys with ALFA:}} There have been three major ``blind''  (untargeted) surveys for 21-cm HI in external galaxies using ALFA away from the Galactic plane. In order of decreasing depth and increasing survey area, these are: the Arecibo Ultra Deep Survey (AUDS), the Arecibo Galaxies Environment Survey (AGES), and the Arecibo Legacy Fast ALFA survey (ALFALFA).\\

\begin{figure*}
\includegraphics[width=0.75\textwidth]{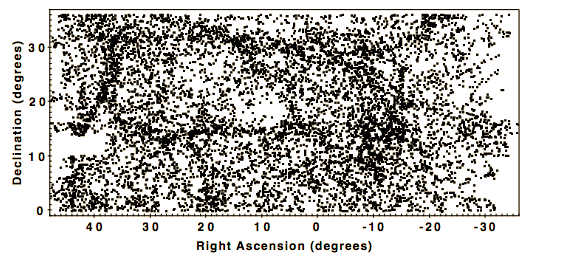}
\caption{The sky distribution of ALFALFA detections in the southern Galactic region. From Haynes et al. (2018) \cite{haynes2018}.}
\label{fig:alfalfa_S} 
\end{figure*}

\emph{{\it ALFALFA}} \cite{alfalfa2005} aimed to map 7000 $deg^{2}$ of the high Galactic latitude sky visible from Arecibo, providing a HI spectral-line database covering a recessional velocity  range of --1600 to +18,000 $km\,s^{-1}$  with ~5 $km\,s^{-1}$ velocity resolution. It covers two sky areas at high Galactic latitude, one in the northern Galactic hemisphere with $07h30m < RA < 16h30m$, and $0^{\circ} < Dec < +36^{\circ}$ and one in the southern hemisphere with $22h < RA < 03h$,  $0^{\circ} < Dec < +36^{\circ}$. Exploiting Arecibo's large collecting area and small beam size, ALFALFA has been  specifically designed to probe the faint end of the HI mass function in the local Universe and provide a census of HI in the surveyed sky area to faint flux limits. The ALFALFA database forms the basis for studies of the dynamics of galaxies within the local and nearby superclusters, allows measurement of the HI diameter function, and enables a first wide-area untargeted search for local HI tidal features. It also searched for HI absorbers at $ z~<~0.06$, and OH megamasers in the redshift range $0.16<z<0.25$. More than 30,000 galaxies were found (many of them new detections) by this survey for which spectra and other parameters are readily available from their data releases 
\footnote{{ https://egg.astro.cornell.edu/index.php/}}. The consortium has been hugely successful with  115 publications between 2005 and 2020. It also mentored generations of students and postdocs, and conducted workshops and other outreach activities. Fig. \ref{fig:alfalfa_S} presents a sky distribution of galaxies in the southern sector from Haynes et al. (2018)\cite{haynes2018}.\\
\emph{{\it AGES}} \cite{auld2006, cortese2008, davies2011} is a deeper survey that concentrates on a number of small regions, covering 200 sq. deg. in total, to 5 times ALFALFA's sensitivity. It is expected to detect $\sim$1000 sources sampling the HI mass function in different environments, and searching for ``dark galaxies''  and  very low-mass objects in nearby groups and clusters. Deshev et al. (2022) \cite{deshev2022} presents the results for one of these fields around Abell 1367,  revealing a continuous correlation of HI-detection fraction (and HI deficiency) with local galaxy density, regardless of the global environment.\\
\emph{{\it AUDS}}, \cite{freud2005} has been designed to make a blind (untargeted) survey of an area of 0.36 square degrees for redshifted HI emission from galaxies beyond the local universe out to a redshift of 0.16 reaching an extremely long on-source integration time of 700 h. The resulting sensitivity  is $\sim$75~$\mu$Jy  per 21.4 kHz, (which is equivalent to 4.5 $km\,s^{-1}$ velocity resolution at z = 0) \cite{hopp2015}. Within the mass range $ log(M_{HI} [h^{-2}\!_{70}] M_{\circ} = 6.32 - 10.76$, 247 galaxies were detected. A value of the cosmic HI density, $\Omega_{HI}\,=\,(3.55~\pm~0.30)\,\times\,10^{-4}~h^{-1}\!_{70}$, was  derived in a 2021 study and showed no evolutionary trend above $2\sigma$ significance between redshifts of 0 and 0.16 \cite{hong2021}.\\
In addition, a fourth, Zone of Avoidance (\emph{\it{ALFA-ZOA}}) survey searched the region behind the Milky Way that is inaccessible at either optical (absorption) or infrared (stellar confusion) wavelengths. The goal has been to find many previously unknown galaxies and reveal large-scale structure behind the plane of the Galaxy \cite{henn2010}. Results from the ``shallow'' pass of the survey has recently been published \cite{szb2019}.\\
\section{SETI at Arecibo} 
As part of the ceremonies marking the dedication of the first telescope upgrade on 16 Nov 1974, Arecibo transmitted a pictorial message at 2.38 GHz aimed at possible inhabitants in globular star cluster M13, $\sim 6.8$~kpc away.  The 1679-bit message included simple images of the telescope, our Solar System, DNA, a human cartoon, and some terrestrial biochemicals. Transmitted over less than 3 min, this was more of an active-SETI demonstration than an attempt at interstellar communication \cite{setimsg1975}.\\
In 1992 the NASA High Resolution Microwave Survey (HRMS) began tracking solar-type stars at Arecibo, searching for signals near 1.4~GHz of extraterrestrial origin. However, after a year its funding was eliminated by Congress. Nevertheless, upon completion of the Gregorian Upgrade, between 1998 and 2004, the SETI Institute privately continued the Targeted SETI search of HRMS at Arecibo, observing in conjunction with the Jodrell Bank 250-ft telescope \cite{backus2002}. This Project Phoenix remained the most sensitive and comprehensive SETI search for almost two decades. Many hundreds of Sun-like stars were observed within 200~ly of the Sun over a band from 1.2 to 3.0 GHz. Almost $2 \times 10^{9} $  frequency channels of 1-Hz width were searched for each star. However, no definitive detection of an extraterrestial intelligence signal was made.\\
The SETI@home project used commensal (piggyback) observations at Arecibo and a huge team of ``volunteer collaborators'' within the general public to search for intelligent signals of extraterrestrial origin
\cite{macrobert1999, anderson2000}. Data were acquired passively in parallel with the main scheduled 
radio-astronomy projects, recorded on storage media, and then transferred to the organizers at the University of California, Berkeley. There they were parsed into small chunks in frequency and time and distributed to the volunteers running Berkeley-developed software on their home computer as screen savers. Results were then communicated back to Berkeley. The project sent out new SETI@home data to volunteers from 1999 to 2020, and involved about 1.8 million volunteer participants. It did not achieve its goal of finding extraterrestrial intelligence, despite identifying several candidate events. However,
a more robust open source software platform  to combat some security breaches of SETI@home was released in 2002. Since then, this Berkeley Open Infrastructure for Network Computing has become an enormously successful backbone for many other applications in areas as diverse as  mathematics,  environmental science, astrophysics, linguistics, medicine, and molecular biology (for example, Rosetta@Home, DENIS@home, asteroid@home, and Einstein@home). SETI@home had also inspired 
the large, hugely successful galaxy detection and classification project, Galaxy Zoo\cite{lintott2008}
in which 100,000 volunteers not only contributed passive computing power, but also actively looked at 
Sloan Digital Sky Survey images taken by a robotic telescope!

\section{Arecibo and VLBI}
VLBI, employing physically unconnected antennas, permits the angular resolution that would be obtainable via a telescope of intercontinental size, or even larger using antennas deployed in Earth's orbit. It uses atomic clocks, usually hydrogen masers, as time and frequency standards, with the data stored on magnetic tape in its earlier days and curently on disc packs, for subsequent transfer to a ``correlation center'' via traditional mail. The correlated data between all participating telescopes in a given VLBI array are then used to create ultra-high angular resolution images of a target radio source. Observational data are sometimes passed directly to the correlation center over the internet, referred to as eVLBI, allowing quasireal time VLBI imaging. While dedicated VLBI arrays have been constructed to exploit the technique, much VLBI is performed with arrays including existing single-dish antennas. The largest single dishes play a special role here in providing the oft-required sensitivity to a particular ad-hoc array.\\
Arecibo participated in the earliest days of VLBI with 610-MHz observation with the Green Bank 140-ft telescope, detecting 12 extragalactic sources, some with sizes $\le 15-20$ mas
\cite{vjaun1970}. In the 1980s occasional VLBI at 18 cm and longer wavelengths was conducted using a MKII recording system. Arecibo also joined a then unprecedented,  28-telescopes Whole Earth Array to image the core of the quasar 3C273 at 1.66 GHz with $\sim$\,3~mas resolution \cite{unwin1994}. \\
Following the Gregorian upgrade, Arecibo reentered the VLBI community seriously, offering not only its immense sensitivity, but also its frequency agility, with achromatic performance up to 10 GHz and a range of state-of-the-art VLBI equipment. Arecibo soon acquired its own VLBA4 tape recording backend, later replacing this with a Mk-5A disc recorder, while JPL loaned a hydrogen maser to the endeavor. Later, RDBE backends, with Mk-5C disc recorders, were implemented. Arecibo was in the process of updating its capabilities to Mk-6 recording when the telescope collapsed in 2020. We note that Arecibo was the sole North American member of the European VLBI Network (EVN), with whom it participated in the first transatlantic eVLBI, with the data correlated in quasireal time in The Netherlands. Later, many such eVLBI runs followed. Arecibo was also a founding member of the High Sensitivity Array (HSA) together with NRAO's VLBA, the phased-up Very Large Array (VLA), the Green Bank Telescope (GBT), and 
the Max Planck Institute for Radio Astronomy’s Effelsberg 100-m dish in Germany.\\
\begin{figure}
 \includegraphics[width=0.5\textwidth]{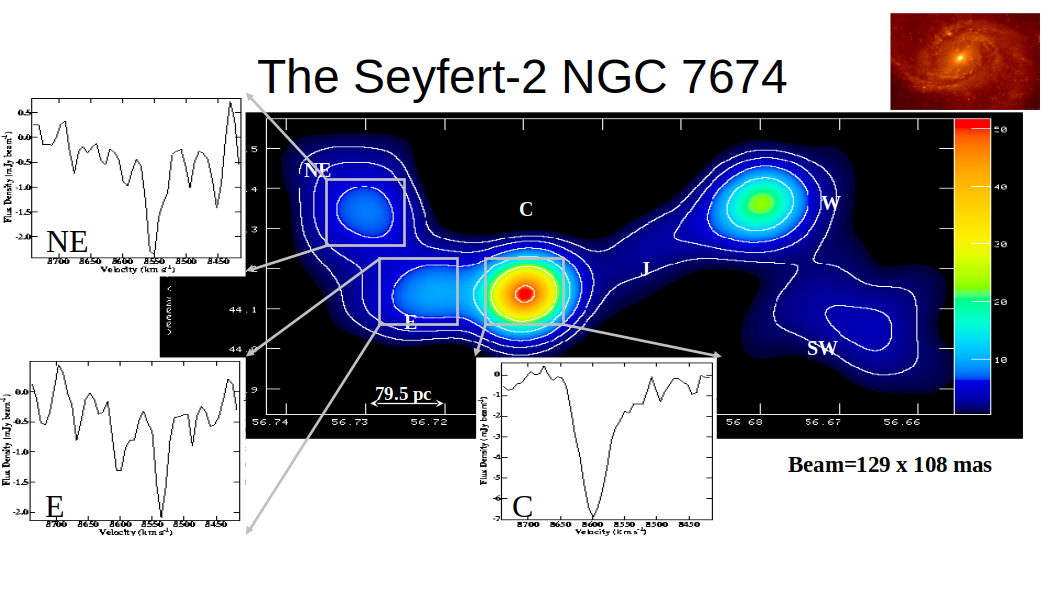}
 \caption{VLBI imaging  of the LIRG, NGC\,7674. From Momjian, 2004 \cite{momjian2003}.}
 \label{fig:n7674}
\end{figure}
\textbf{\emph{AGNs in ULIRGs:}} Soon after the implementation of the VLBA4 system at Arecibo, a project 
that demonstrated the power of including the 305-m dish in a VLBI array was performed by Momjian et al. (2003) \cite{momjian2003}. This made phase-referenced observations of both the radio continuum and $\lambda\,21$-cm HI absorption toward the luminous infra-red galaxy (LIRG), NGC~7674. This used an array consisting of Arecibo, the VLBA, and the phased VLA. With a variety of resolutions, it revealed continuum structure related to activity of an AGN; see Fig. \ref{fig:n7674}. The VLBI HI-absorption imaging enabled inference of the presence of an inclined HI disc or torus feeding the central AGN. The authors noted that with Arecibo present for 20\% of the observing time, the image sensitivity was $\sim$4.2 times better than with the VLBA alone for the entire time.\\ 
\textbf{\emph{The Pleiades Distance Controversy:}} Also demonstrating the great ability of 
the upgraded Arecibo telescope to contribute  to terrestrial VLBI up to 10 GHz was a project 
of Melis et al. (2014) \cite{melis2014}. This targeted a number of young stars in the nearby 
Pleiades open cluster, whose distance is a cornerstone in the ``cosmic distance ladder''. 
A conflict \cite{narayanan1999} had arisen  between the distances to the cluster determined 
from the Hipparcos space-astrometry mission in 1997 (i.e. 120.2 ± 1.5 pc), and that derived 
via several ground-based techniques (133.5 ± 1.2 pc). 
The Pleiades is an open cluster made of young stars that are only about 100 million years old.
Their distances were measured using the apparent brightness of the stars compared
with their intrinsic brightness based on models of their stellar burning processes.
If the Hipparcos distance is correct, it challenges scientists’ understanding of young stars
— and everything that happens to them afterward, including the fate of their planetary 
systems.
Via 8.4-GHz VLBI measurements to a precision of better than 100 $\mu$as, Melis et al. (2014) determined the trigonometric parallaxes of four Pleiades stars directly in the International Celestial Reference Frame, ICRF(S/X)  defined by distant quasars. Without the presence of Arecibo's great sensitivity, following the weak radio continuum from these stars  with VLBI's spatial resolution would have been impossible. As these were also in binary systems,  Melis et al. monitored their precise locations weekly, over $\sim$~1.5 years, to solve both their orbits and their parallaxes  and derived the distance 
to the cluster to be 136.2 ± 1.2 pc. This value has later been confirmed by measurements using Gaia DR1 and DR2 (Gaia Collaboration  2016 and 2018 \cite{gaiadr2016a, gaiadr2016,gaiadr2018}). The 
reason for the long-standing discrepancies between the Hipparcos and the ground-based measurements has been the subject of much discussion since 2002 (e.g \cite{makarov2002, makarov2022}. Ultimately, the distance controversy has been settled in favor of the ground-based estimates, with Arecibo contributing to this foundational work by providing its VLBI capabilities to the measurements with the smallest error bars; see Fig .\ref{fig:pleiades}.\\
\begin{figure}
 \includegraphics[width=0.5\textwidth]{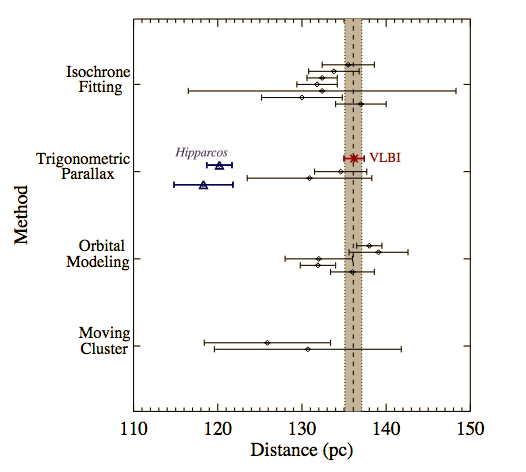}
 \caption{Summary of Pleiades distances obtained through various methods. The red asterisk with a distance of 136.2±1.2 pc is the VLBI determination of Melis et al \cite{melis2014}. }
 \label{fig:pleiades}
\end{figure}

\begin{figure}
 \includegraphics[width=0.5\textwidth]{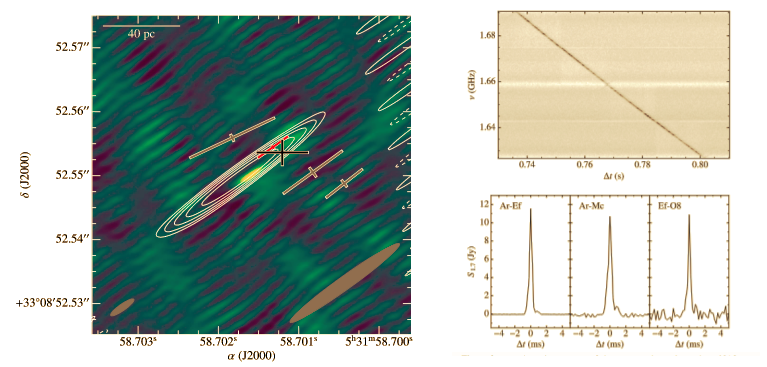}
 \caption{{\bf Left:} EVN image of the persistent source at 1.7 GHz (white contours) associated with the repeating FRB\,121102. The positions of the strongest burst (red cross), the other three bursts (gray crosses), and the average of all four bursts detected on 2016 September 20 (black cross) are also marked. {\bf Right, Top:} Dynamic spectrum of the strongest burst showing the dispersive sweep across the band. {\bf Right, Bottom:} coherently dedispersed and band-integrated profiles of the same burst seen in the cross-correlations for Arecibo–Effelsberg (Ar-Ef), Arecibo–Medicina (Ar-Mc), and Effelsberg–Onsala (Ef-ON) baselines. From Marcote et al., 2017 \cite{marcote2017}. }
 \label{fig:frbar}
\end{figure}

\textbf{\emph{Host galaxy of the repeating fast radio burs  FRB\,121102:}}
In Sept. 2016 Arecibo played a major part with the EVN in making the first VLBI imaging at 1.7~GHz of four individual millisec-duration bursts from a {\emph{fast radio burst (FRB)}}, the first ``repeating FRB'', FRB\,121102.\\
FRBs are the latest arrival among the enigmatic objects in the transient radio sky. Discovered in 2007 [the Lorimar Burst \cite{lorimer2007}] these extremely short-duration (typically about a millisecond), highly energetic radio bursts -- which are often just a single burst -- have now (at the time of writing) been detected in about 1000 locations all across the sky. 
Their high dispersion measures imply cosmological distances and the nature of the bursts point toward cataclysmic events involving highly magnetized neutron stars, or black holes. 
About 10\% of these bursts have been observed to be repeating, with the first
ever ``repeating FRB'' detected at Arecibo in 2014 \cite{spitler2014}. A recent review of the 
FRB research provides the current status of this exciting field \cite{lorimer2024}. 
Although, owing to its limited sky coverage and relatively slow slewing rate, Arecibo had not been the telescope of choice for FRB discoveries, its sensitivity proved vital in the first ever direct detection 
of the host galaxy of a FRB. Arecibo served as the ``burst detector'', in parallel 
providing data to the EVN via eVLBI in a 2016 observing run. Short stretches of VLBI data were imaged for the four bursts detected by Arecibo, and all were detected by VLBI. In addition, the complete database was imaged to map the ``persistent source'' previously found by the VLA to exist in the host galaxy of the FRB. The FRB and the persistent source were found to be colocated (see Fig. \ref{fig:frbar}) to $\lesssim$40\,pc, (i.e.  $\lesssim$12\,mas). The investigators'  preferred interpretation for FRB\,121102 was that it either is associated with a low-luminosity galactic nucleus, or is a young neutron star energizing a supernova remnant \cite{marcote2017}.\\

\textbf{\emph{Arecibo in Space VLBI:}}
Arecibo's immense collecting area became particularly important in its collaboration with two relatively small VLBI antennas in space. These were, (a) at 1.6 and 5 GHz with the Japanese VSOP/HALCA 8-m diameter telescope, and (b) with the Russian 10-m diameter dish of the {\emph{RadioAstron} mission. While the apogee of HALCA's orbit was $\sim$2~Earth diameters, RadioAstron's apogee was $\sim$30~Earth diameters, almost the distance to the Moon. For the observations with HALCA, made soon after the Gregorian upgrade, Arecibo was loaned a Canadian S2 VLBI recorder. A highlight was the 5-GHz Deep Interferometric VSOP--Arecibo Survey (Edwards et al. 2009 \cite{edwards2009}).\\ 
Arecibo was a major player in collaboration with {\emph{RadioAstron}}, which launched in 2011, participating in over 400 individual single-frequency observing sessions at P, L or C bands. The results from the observations of pulsar PSR B0950+08 have contributed to an improved understanding of radio-wave scattering by the interstellar plasma \cite{smirnova2014}. The observations
by Kovalev et al. (2016) \cite{kova2016} challenged the brightness-temperature limit set by inverse Compton cooling for the quasar 3C273. A likely explanation was provided through refractive substructure
by Johnson et al (2016) \cite{johns2016}. Major conclusions from the {\emph{RadioAstron}} Key Science survey of AGNs were presented by Kovalev et al.\cite{kov2020}.\\
\section{Other Recent Arecibo Results}
\textbf{\emph{Individual pulsar studies:}} These have yielded hitherto unknown nanosecond substructures in the occasional giant pulses emitted by the Crab pulsar, indicating beach-ball-sized emission regions in the magnetosphere of this pulsar \cite{hankins2003Nat}. The {\emph{strong equivalence principle}} has been investigated using  pulsar timing \cite{freire2003}, while ``new'' interstellar scattering effects have also been uncovered at Arecibo \cite{stineb2007}. High-quality polarimetric pulsar measurements have been made at Arecibo for more than 40 years (e.g. \cite{stineb1984}). In particular, a long series of such observations associated with Rankin and collaborators have enabled interpretation in terms of the associated emission-beam structure; see Rankin et al. (2023), and Wahl et al. (2013) \cite{rankin2023, wahl2023} and references therein).\\

\textbf{\emph{The Coolest Radio-flaring Brown Dwarf:}}
Radio detection provides a unique means to measure and study magnetic fields of the coolest brown dwarf stars. Previous radio surveys had observed quiescent and flaring emission from brown dwarfs (due mostly to the electron cyclotron maser instability) down to spectral type L3.5, but only upper limits had been established for even cooler objects. Route \& Wolszczan (2012) \cite{route2012} reported the detection of sporadic, circularly polarized flares from the T6.5 dwarf 2MASS J1047+21, with the Arecibo radio telescope at 4.75 GHz. This was by far the coolest brown dwarf yet detected at radio frequencies. The fact that such an object is capable of generating observable, coherent radio emission, despite its low, 900~K temperature, demonstrates the presence of substellar magnetic activity and the feasibility of studies of brown dwarfs of spectral types {\emph{L, T,}} and {\emph{Y}}, using radio detection as a tool. \\

\textbf{\emph{Extragalactic ``prebiotic'' molecules:}}
Using a then-unprecedented 800-MHz bandwidth spectrometer, a dozen molecular lines were found by Arecibo in the ULIRG Arp~220. These included the first distant extragalactic detection of the prebiotic molecule Methanimine ($CH_{2}NH$) \cite{salter2008}; see Fig. \ref{fig:ch2nh}).  Methanimine is considered to be a ``prebiotic'' molecule because a combination with Hydrogen Cyanide (HCN; also detected in Arp220 at Arecibo) followed by a process of hydrolysis, can produce Glycine, the simplest of the 20  amino acids that constitute all proteins.\\
\begin{figure}
 \includegraphics[width=0.5\textwidth]{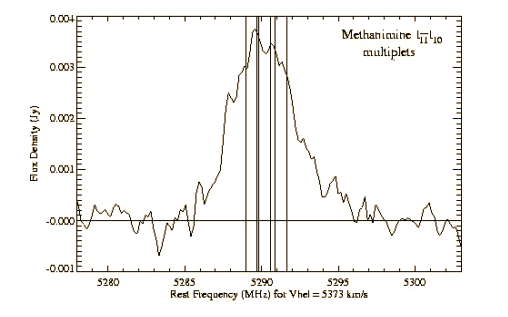}
 \caption{The first distant extragalactic detection of the prebiotic molecule, Methanimine, $CH_{2}NH$, in Arp220. From Salter et al., 2008. \cite{salter2008}}
 \label{fig:ch2nh}
\end{figure}

\textbf{\emph{The most stringent constraints yet on the variation of fundamental constants:}} 
\begin{figure*}
 \includegraphics[width=0.75\textwidth]{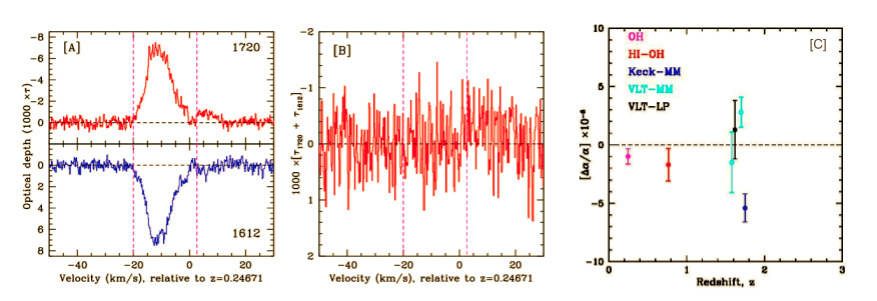}
 \caption{[{\bf a}]: OH satellite lines in conjugation. [{\bf b}]: Difference of the velocity structures of the satellite lines. [{\bf c}]: Limits on the variation of the fine-structure constant. From Kanekar, Ghosh, and Chengalur, 2018 \cite{kanekar2018}.} 
 \label{fig:fc2017}
\end{figure*}
Comparisons between the redshifts of multiple spectral transitions in distant galaxies provide a sensitive probe of evolution in fundamental constants such as the {\it fine-structure constant ($\alpha$) and the proton-electron mass ratio ($\mu$)} over cosmological epochs. The rotational ground state of the OH radical is split into four sublevels by lambda doubling and hyperfine splitting. In certain astrophysical circumstances, the satellite OH 18-cm lines have the same shape, but with one line in emission and the other in absorption. This “conjugate” behavior arises owing to an inversion of the level populations within the OH ground state (\cite{vanLan1995}). Only a single completely conjugate satellite OH system is currently known, at $z \sim 0.247$ toward PKS 1413+135 \cite{kanek2004, darling2004}. Using the Arecibo Telescope, Kanekar et al. (2018) \cite{kanekar2018} carried out one of the deepest-ever integrations (140 hr) in radio spectroscopy, targeting the redshifted conjugate satellite OH 18-cm lines toward PKS~1413+135. Their  analysis yielded $\Delta X/X$~=~$(-1.0~\pm~1.3)~\times~10^{-6}$, which is consistent with no changes in the quantity, $ X~=~\mu~\alpha^{2}$ over the last 2.9 Gyr. With the reasonable assumption that $\Delta\alpha/\alpha >> \Delta \mu/\mu$, this asserts the most stringent constraint on fractional changes in the fine-structure constant from astronomical spectroscopy.\\

In the optical regime, the use of echelle spectrographs on 10-m class optical telescopes, and the many-multiplet method \cite{dzuba1999}, based on rest-frame ultraviolet spectral lines, has yielded the
highest sensitivity to changes in $\alpha$ out to much higher redshifts \cite{murphy2016, kotu2017}. 
As discussed by Kanekar et al. (2018) \cite{kanekar2018}, while the optical results have low statistical errors, it has also become clear that most of these studies are afflicted by systematic errors (see \cite{griest2010, whitmore2015}) and still contradict each other. 
The use of the 18-cm OH conjugate lines  allows one to measure changes in the constants from a single space-time location without the need to average over multiple absorbers (as would be required in most other techniques to overcome possible local velocity offsets between the gas clouds giving rise to the different lines).
Figure \ref{fig:fc2017} demonstrates the complementary view on the evolution of the fine-
structure constant that optical and radio observations provide, as well as the most stringent nature of the Arecibo estimation of it.\\
\begin{figure}
 \includegraphics[width=0.475\textwidth]{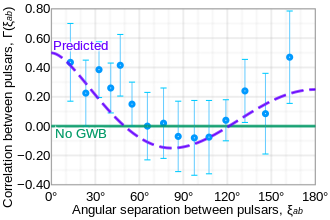}
 \caption{Plot of correlation between pulsars observed by NANOGrav vs angular separation between the pulsars, compared to a theoretical model (or Hellings–Downs curve; dashed purple) and if there were no gravitational wave background (solid green). From Agazie et al., 2023 \cite{agazie2023}. } 
 \label{fig:gw2023}
\end{figure}

\textbf{\emph{Stochastic Background Gravitational Wave detection using millisecond pulsars; the 2023 NANOGrav result:}}  Fifteen years of meticulous and painstaking work accurately timing millisecond pulsars recently produced the first-ever direct detection of a stochastic gravitational-wave background (see Fig.\ref{fig:gw2023} \cite{agazie2023}). The method relies on the effect of a passing gravitational wave to perturb the local space-time and cause a change in the observed rotational frequency of a pulsar. For an array of pulsars distributed at different distances and position angles from the observer, Hellings and Downs \cite{heldow1983} showed that a stochastic background of gravitational waves would produce a correlated signal for different angular separations on the sky. Using accurate, regular measurement of pulse arrival times between 2004 and 2019 at Arecibo and the GBT (and also the phased VLA for part of the time), the NANOGrav  consortium have now detected this effect \cite{agazie2023}. The formation of the consortium itself was the result of the great attraction of Arecibo to generations of pulsar observers. There have been similar international projects such as the International Pulsar Timing Array, the European Pulsar Timing Array, and the Chinese Pulsar Timing Array. Together, these represent a large worldwide collaboration and are contributing to ``multi-messenger'' astronomy.
\section{Fifty-seven Years At The Cutting-Edge}
Although the Arecibo 305-m telescope served radio astronomers for 57 years, it never became obsolete 
as a frontline research instrument. This was possible only through the three major upgrades that the telescope underwent during those years. In its own way, each upgrade produced a new telescope for its 
users, offering state-of-the-art performance whenever its existing configuration might have otherwise appeared near to being fully exploited. The most important consideration of its three upgrades was to maximize the parameter space that could be exploited by potential users.\\
For the majority of its life, the 305-m telescope was a national facility, open to users globally. 
Over more than the final two decades of its life, this was achieved through regular proposal calls, with 
the proposals evaluated via anonymous referees, and observing time allocated according to these 
evaluations. Time on the telescope was always heavily over-subscribed, right up until its collapse 
in 2020. \\
Originally conceived as a tool for upper atmospheric research, Professor William Gordon foresaw 
that the instrument offered unique potential both for studies of planetary physics via its powerful 
meter-wavelength radar, and for low-frequency radio astronomy. A decade later, the first
Arecibo upgrade was designed to permit observations at short decimeter wavelengths. The major 
beneficiaries of this were the astronomical community, both planetary [see the companion
paper by Nolan, Carter, and Rivera-Valentín (2026) \cite{nolan2026planetary}] and extra-Solar System observers.\\ 
The proposed expansion of the telescope's parameter space for the upgrade of the mid-1990s was 
evaluated in detail. A consideration of the potential offered by the planned expansion of the upper frequency, plus the improved sensitivity, and provision of achromatic performance was considered at the  Arecibo Upgrading Workshop in October 1986, which attracted nearly 50 participants. This resulted in a 
proceedings in which an anticipated scientific program for the upgraded telescope was set out. 
It is notable that most of the suggested areas of research then discussed later appeared within the instrument's actual program. The greatly increased parameter space of the first two 
upgrades also led to a happy number of serendipitous discoveries. 
The desire of many investigators to use the post-2000 telescope for large-area sky surveys resulted directly to the ALFA upgrade a decade later; the increase of instantaneous sky coverage by a factor 
of 7 made this a much more attractive proposition.  These large-area surveys were not reasonable subjects for standard telescope proposals, as they involved an order of magnitude more observing time 
than even the larger proposals received up to that time.  Arecibo Observatory resolved this issue by 
inviting all interested scientists to join one of a number of consortia that aimed to exploit the new survey capabilities. This worked out well, with the individual multiscientist consortia deciding their 
detailed observing strategies, and even cooperating across consortia boundaries to exploit joint, (commensal) observing. The successes of a number of these consortia are summarized in 
Sec. V.B.\\
Despite its unusual appearance, the 305-m telescope was essentially a standard single-dish 
instrument and hence was responsive to the shortest spatial frequencies of the distribution of 
celestial radiation. This aspect of its parameter space was vital to the complete Arecibo sky
GALFA-HI and GALFACTS ALFA surveys; see Sec. V.B. The single-dish aspect of Arecibo also 
facilitated the observatory's enthusiasm for potential users to bring their personal backend 
equipment when this could provide possibilities not covered by the generalized processors
provided by Aecibo Observatory. A number of such investigators generously made their equipment 
available to the global user base, often in exchange for servicing and maintenance by the observatory engineering staff. 
As for all single dishes, the 305-m telescope engaged in a constant struggle with radio
frequency interference. A major achievement in this endeavor was the establishment in 1997 
of the Puerto Rico Coordination Zone (PRCZ) by the Federal Communications Commission to 
protect the observatory site as much as possible from harmful interference, up to 15 GHz, 
from Puerto Rico and its associated islands \footnote{ See {https://www.fcc.gov/document/radio-astronomy-coordination-zone-puerto-rico}}. The PRCZ is still being operated under the management of NRAO.\\
\section{Concluding Remarks} 
As an instrument with a chicken-wire surface, costing just under 10 million US dollars in 1963, 
with radio astronomy a minor interest, the Arecibo 305-m telescope then remained 
on the radio-astronomy cutting edge for the next 57 years. Through those years it expanded the 
range of possibilities it offered its users, retaining its place as a primary astronomical research instrument to the end.
At the time of its collapse, Arecibo was due to install an ultrawideband receiver and a 40-beam, phased-array feed that would have resulted in yet another quantum leap in its capabilities. Nearly 60 years after its first light, the telescope's demise still felt premature. What is left behind is the remarkable legacy of its achievements, large archives of data still to be fully processed, and many astronomers worldwide who learned their trade at this little corner of the ``Isla del Encanto''.\\
For a full history of the Arecibo Observatory, from its genesis in 1958 to its closure in
2023, we highly recommend the recently published book by Campbell (2025)
\cite{campbell2025arecibo}.
\section*{Acknowledgements}
The Arecibo 305-m telescope was part of the National Astronomy and Ionosphere Center. Under cooperative agreements with the National Science Foundation, it was operated by Cornell University until 2011, by SRI International and associates between 2011 and 2018, and by a consortium led by the University of Central Florida from 2018 to the closure of the observatory in 2023. The authors are grateful for having been a part of the broad “Arecibo family” of staff, students, and visiting scientists for over a quarter of a century.
\bibliography{reference}
\end{document}